\documentclass{isprs}
\usepackage{subcaption}
\usepackage{graphicx, pgfplots}
\usepackage{setspace}
\usepackage{tikz}
\usepackage{MnSymbol,utfsym}
\usepackage{arydshln}
\usepackage{tabularx,booktabs}
\usepackage{fancyhdr}
\usepackage{geometry} 
\usepackage[%
  breaklinks=true,
  letterpaper=true,
  colorlinks,
  bookmarks=false,
]{hyperref}
\hypersetup{linkcolor={black}, urlcolor={blue}, citecolor={black}, colorlinks=true}

\usepackage[capitalize]{cleveref} 
\usepackage{microtype}

\usepackage[%
  xindy,
  acronym,
]{glossaries}
\glsdisablehyper

\usepackage{natbib}

\usepackage{siunitx}

\newacronym{gl:DSM}{DSM}{digital surface model}
\newacronym{gl:DEM}{DEM}{digital elevation model}
\newacronym{gl:SR}{SR}{super-resolution}
\newacronym{gl:LiDAR}{LiDAR}{light detection and ranging}
\newacronym{gl:InSAR}{InSAR}{interferometric synthetic
aperture radar}
\newacronym{gl:SISR}{SISR}{single image super-resolution}
\newacronym{gl:GDSR}{GDSR}{guided depth map super-resolution}
\newacronym{gl:GAN}{GAN}{generative adversarial network}
\newacronym{gl:DKN}{DKN}{deformable kernel networks}
\newacronym{gl:CNN}{CNN}{convolutional neural network}
\newacronym{gl:GSD}{GSD}{ground sampling distance}

\newcommand{\RealGDSRdot}{Real-GDSR}
\newcommand{\RealGDSR}{\RealGDSRdot\space}

\renewcommand{\headrule}{} 

\begin{document}

\title{ Real-GDSR: Real-World Guided DSM Super-Resolution via Edge-Enhancing Residual Network}

\author{
 Daniel Panangian\thanks{Corresponding author}
 , Ksenia Bittner}


\address
{
	Remote Sensing Technology Institute, German Aerospace Center (DLR), Wessling, Germany \\(daniel.panangian, ksenia.bittner)@dlr.de
}

\abstract
{A low-resolution \gls{gl:DSM} features distinctive attributes impacted by noise, sensor limitations and data acquisition conditions, which failed to be replicated using simple interpolation methods like bicubic. This causes super-resolution models trained on synthetic data does not perform effectively on real ones. Training a model on real low and high resolution \glspl{gl:DSM} pairs is also a challenge because of the lack of information. On the other hand, the existence of other imaging modalities of the same scene can be used to enrich the information needed for large-scale super-resolution. In this work, we introduce a novel methodology to address the intricacies of real-world \gls{gl:DSM} super-resolution, named \textsc{\RealGDSRdot}, breaking down this ill-posed problem into two steps. The first step involves the utilization of a residual local refinement network. This strategic approach departs from conventional methods that trained to directly predict height values instead of the differences (residuals) and utilize large receptive fields in their networks. The second step introduces a diffusion-based technique that enhances the results on a global scale, with a primary focus on smoothing and edge preservation. Our experiments underscore the effectiveness of the proposed method. We conduct a comprehensive evaluation, comparing it to recent state-of-the-art techniques in the domain of real-world \gls{gl:DSM} \gls{gl:SR}. Our approach consistently outperforms these existing methods, as evidenced through qualitative and quantitative assessments.
}


\maketitle


\glsresetall

\section{Introduction}\label{MANUSCRIPT}

Elevation data are essential for a wide range of applications across multiple sectors. They also contribute significantly to our improved understanding and management of the Earth's resources and environment. These data are collected through various techniques, typically aerial data capture and \gls{gl:LiDAR} sensors. A \gls{gl:DEM} is a digital representation of the Earth's topography, specifically focusing on the bare ground surface. \Glspl{gl:DEM} play a crucial role in applications like terrain analysis, hydrological modeling, geological studies, precision agriculture, and infrastructure planning. A \gls{gl:DSM} in the other hand is a representation which includes all objects on it, such as vegetation and buildings. \Glspl{gl:DSM} are widely applied in topographic mapping, environmental simulations \citep{Aktaruzzaman2009}, 3D city modeling and planning. Recent remote sensing technology provides several ways to measure the 3D urban morphology. Conventional ground surveying, stereo airborne or satellite photogrammetry, \gls{gl:InSAR}, and \gls{gl:LiDAR} are the main data sources used to obtain high-resolution \gls{gl:DSM}. However, each of these scenarios has its own set of advantages and disadvantages. 

\begin{figure}[t]
\begin{center}
\includegraphics[width=8cm]{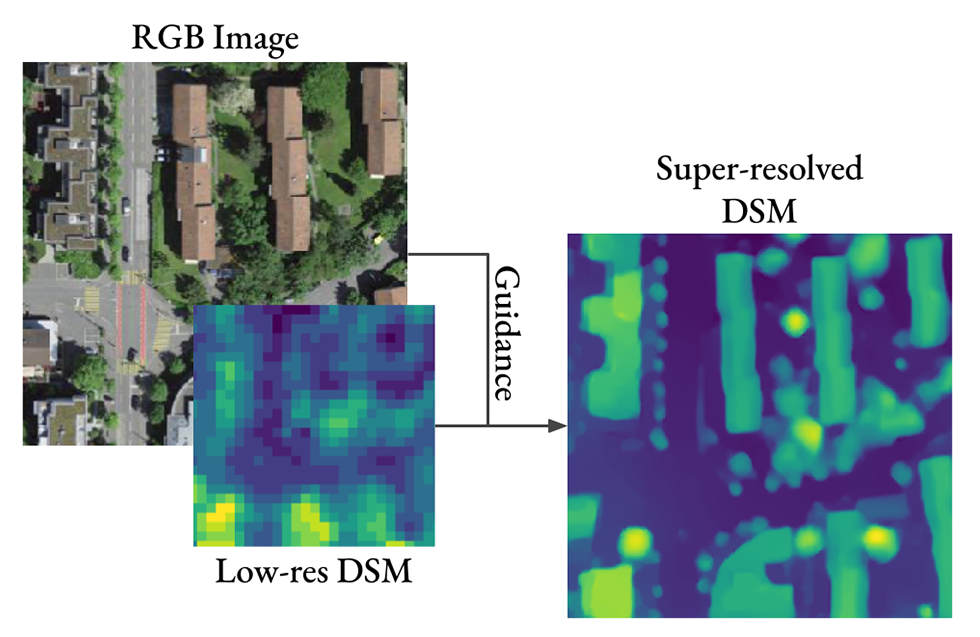}
\end{center}
   \caption{Guided \gls{gl:DSM} super-resolution: given a low-resolution
dsm and a high-resolution guide image, our method
predicts a high-resolution \gls{gl:DSM}. The figure shows an
example output of the proposed method on low-resolution \gls{gl:DSM} with a factor of 10.}
\label{fig:overview}
\end{figure}

Data derived from terrestrial and airborne systems offer high spatial resolution but have limited coverage and can encounter precision-related issues. Spaceborne missions, for example Cartosat 1, provide global coverage but may lack the finest level of resolution achieved by terrestrial and airborne methods due to challenges in capturing high-resolution data from space. The resolution of such data have a substantial impact in different fields of operation. Improving measurement equipment precision is the most straightforward way to acquire high-resolution elevation data, but it is a difficult, costly, and time-consuming procedure. Therefore, generating high-resolution data without extra cost becomes a key concern of researchers from various fields. A practical way to solve this problem is to enhance the resolution of easily obtained low-resolution data~\citep{Xu2015}. In computer vision, various algorithms have been developed for single image super-resolution task. Some of the algorithms have been explored for \gls{gl:DEM} super-resolution. In addition, The depth estimation ~\citep{Godard2017,Godard2019,Bhat2021} and depth super-resolution task ~\citep{Voynov2018,He2021} are two intriguing applications that can be applied to DEMs and DSMs. Recently a method termed guided depth map super-resolution \gls{gl:GDSR} is gaining popularity. The idea is that a different imaging modality of the same object can be used as a guide for super-resolving the low-resolution image by injecting the missing high-frequency content. Research into guided depth super-resolution has a long history~\citep{Patterson1992,Izraelevitz1994}. The proposed solutions range from classical, entirely hand-crafted schemes~\citep{Ham2017} to fully learning-based methods~\citep{Hui2016}, while some recent works have combined the two with promising results~\citep{Lutio2022,Metzger2023}. 

\begin{figure}
    \begin{subfigure}{\linewidth}
    \centering
        \begin{subfigure}{.32\linewidth}
            \includegraphics[width=\linewidth]{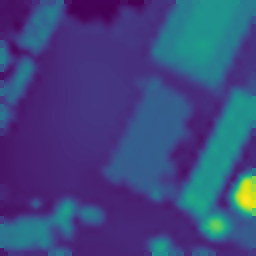}
        \end{subfigure}
        \begin{subfigure}{.32\linewidth}
            \includegraphics[width=\linewidth]{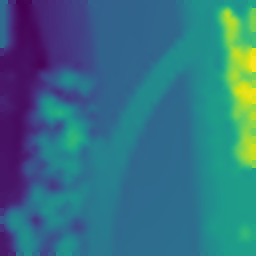}
        \end{subfigure} 
        \begin{subfigure}{.32\linewidth}
            \includegraphics[width=\linewidth]{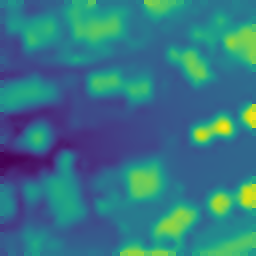}
        \end{subfigure}
        \caption{Bicubic Downsampled}
    \end{subfigure}
    \begin{subfigure}{\linewidth}
        \centering
        \begin{subfigure}{.32\linewidth}
            \includegraphics[width=\linewidth]{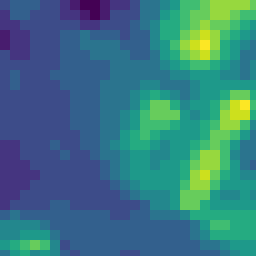}
        \end{subfigure}
        \begin{subfigure}{.32\linewidth}
            \includegraphics[width=\linewidth]{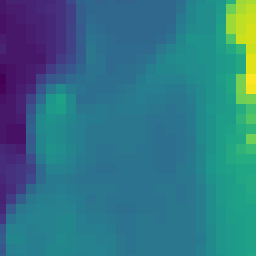}
        \end{subfigure}
        \begin{subfigure}{.32\linewidth}
            \includegraphics[width=\linewidth]{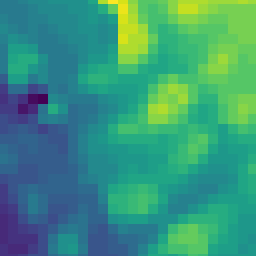}
        \end{subfigure}
        \caption{Real Low Resolution}
    \end{subfigure} 
        \begin{subfigure}{\linewidth}
        \centering
        \begin{subfigure}{.32\linewidth}
            \includegraphics[width=\linewidth]{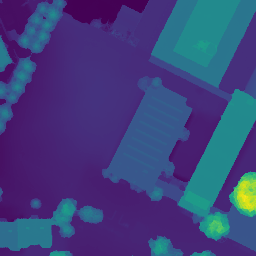}
        \end{subfigure}
        \begin{subfigure}{.32\linewidth}
            \includegraphics[width=\linewidth]{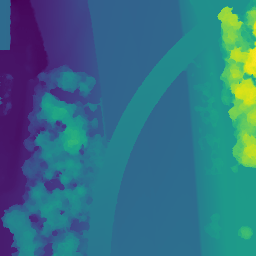}
        \end{subfigure}
        \begin{subfigure}{.32\linewidth}
            \includegraphics[width=\linewidth]{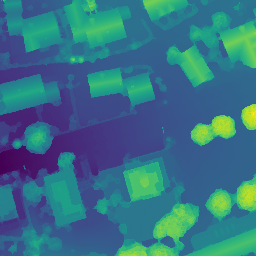}
        \end{subfigure}
        \caption{High Resolution}
    \end{subfigure} 
    \caption{Examples of real low-resolution \glspl{gl:DSM} in comparison to the bicubic-downsampled and high-resolution \glspl{gl:DSM}. Note that real low-resolution \glspl{gl:DSM} preserve less information in comparison to their bicubic-dowsampled counterparts.}
    \label{fig:real-low-dsm}
\end{figure}

However, generally both \gls{gl:SISR} and \gls{gl:GDSR} methods restrict themselves to super-resolving images downsampled by a simple and uniform degradations (i.e, bicubic downsampling). Real low-resolution \glspl{gl:DSM}, in the other hand, do not preserve as much information (see \cref{fig:real-low-dsm}) and applying these methods on real-world \glspl{gl:DSM} is becoming a challenge ~\citep{cai2019toward, wang2021real}. In this paper, we focus on super-resolving \glspl{gl:DSM} guided by their corresponding optical images, which is depicted in ~\cref{fig:overview}. We focus our research on urban \glspl{gl:DSM} because they contain richer information which hard to be restored. This focus also aligns with the evolving interest in leveraging DSMs for building reconstruction \citep{bittner2018dsm,partovi2019automatic,bittner2020long,wang2021machine,gui2021automated,stucker2022resdepth,stucker2022implicity}. 

\pagestyle{fancy}
\fancyhf{} 
\fancyfoot[C]{\thepage} 
\renewcommand{\headrule}{} 

In summary, our contributions are as follows:
\begin{enumerate}
  \item We propose a novel super-resolution framework guided by optical images on real-world low-resolution \glspl{gl:DSM}. To the best of our knowledge, we are the first to develop a guided super-resolution network trained on real-world DSMs.
  
  \item We achieve this by utilizing both local and global context. We improve the conventional methods in two ways: local refinement and edge-enhancing diffusion. A local refinement network with small receptive field is utilized to improve the low-resolution \gls{gl:DSM}. This shallow network can repair some missing regions and structures, according to the surrounding local regions after a coarse bicubic-interpolation stage. An edge-enhancing diffusion network is used to further smoothen the super-resolved result. This network can further improve the visual quality using the global edge information, especially for removing outliers and preserving height discontinuities.
  
    \item We demonstrate the effectiveness of our approach by comparing it to other state-of-the-art networks, trained in the same setting. Furthermore, we show that our approach is feasible of achieving a 10x super-resolution for low-resolution \glspl{gl:DSM} from satellite data. We also found that our proposed approach outperformed the other works in terms of both qualitative and quantitative results on unseen data.
\end{enumerate}

\section{Related Work}\label{sec:TITLE AND ABSTRACT BLOCK}

\subsection{Single Image Super-Resolution}\label{sec:Title}

Single image super-resolution refers to generation of a high-resolution image from a low-resolution image. Recently \gls{gl:SISR} methods has shifted towards example-based approaches. However in those early years interpolation-based techniques like linear, bicubic or Lanczos are widely used. The idea is new pixels are estimated by interpolating given pixels. But these methods suffers from blurry results on high-frequency regions. Learning or example-based methods aim to gather insight information from paired low and high-resolution images to understand missing details in low-resolution images. SRCNN \citep{dong2015image} is one of the first approaches to demonstrate the use of neural networks to learn the nonlinear mapping of the images in the image space. The method utilize bicubic interpolation of low-resolution image followed by high-dimensional vector representation and ended by reconstruction of the vectors to the pixel space. VSDR \citep{kim2016accurate} was designed to increase the efficiency of SRCNN, by predicting the residuals rather than the actual pixel values and, to boost the overall performance by adding more layers. \Citet{wang2015deep} introduced sparse coding to the training which enable the model to enlarge the images to the desired scale factor progressively. Deep recursive layers are introduced in DRCN \citep{kim2016deeply}  to reduce the number of parameters.  

Furthermore, the introduction of \glspl{gl:GAN} inspired the implementation of \gls{gl:GAN} for super-resolution. In \gls{gl:GAN} the generator produces high-resolution image and the discriminator is trained to distinguish between
patch of the original image and patch which are produced by the generator. Two components are designed to defeat each other in a zero-sum game \citep{goodfellow2014generative}. SRGAN \citep{ledig2017photo} is the first one to demonstrate the ability to pay more attention to visual effects, introducing adversarial and perceptual losses. ESRGAN \citep{wang2018esrgan} goes another step by introducing relativistic discriminator and removing batch normalization layers.

\begin{figure*}[ht]
\begin{center}
\includegraphics[width=16cm]{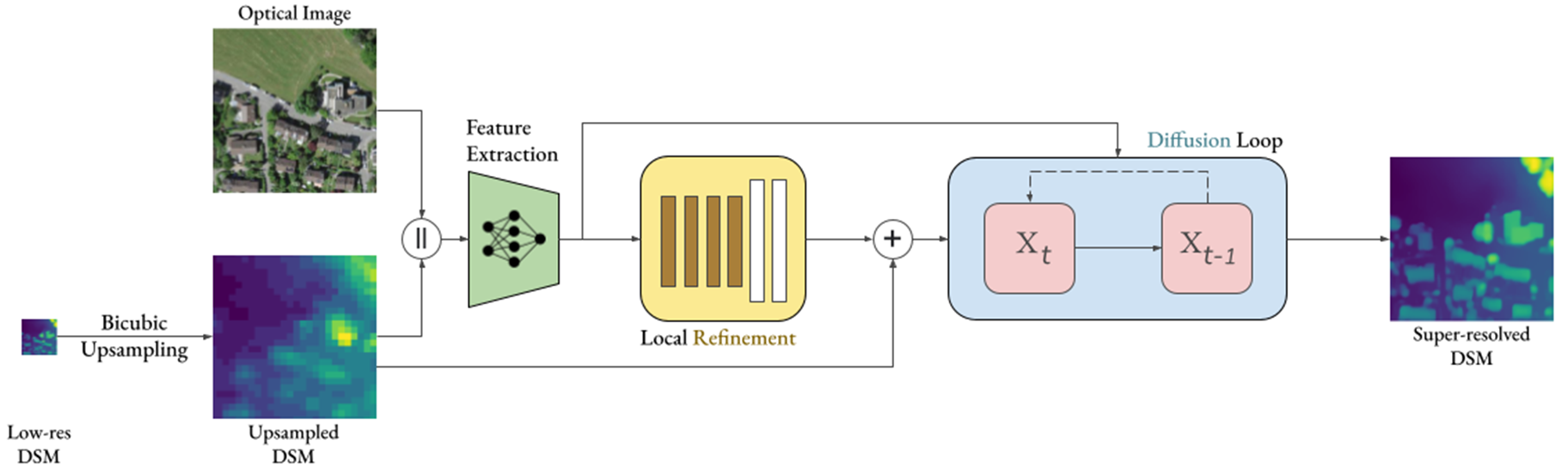}
\end{center}
   \caption{Summary of the proposed architecture. Real-GDSR comprises mainly a two-step process: Initially, high-dimensional features are extracted from both bicubic-upsampled low-resolution DSM and high-resolution optical image by a pre-trained model. Subsequently, a local refinement network refines the upsampled DSM by incorporating residual blocks and upsampling operations, followed by a diffusion network, which iteratively enhancing the refined upsampled DSMs, emphasizing edge features from the high-resolution optical image.}
\label{fig:realgdsr}
\end{figure*}

 But different than images, depth images contain piecewise affine regions that have sharp depth discontinuities and no textures \citep{riegler2016atgv}. Besides, they are sensitive to artifacts \citep{xie2015edge}. In this paper, we focus on the training strategy and framework of super-resolution networks when dealing with depth images.

\subsection{Guided Depth Map Super-Resolution}


\Gls{gl:GDSR} has become an essential topic in multi-modal image processing and super-resolution. The idea behind it is that there are statistical co-occurrences between the texture edges of RGB images and the discontinuities of depth maps \citep{xie2015edge}. Hence, information in RGB images can be utilized to restore low-resolution depth maps. Recently conventional methods for \gls{gl:GDSR} can be divided into three categories: learning-free, learning-based and hybrid approaches. Early work on \gls{gl:GDSR} consisted mostly of filter-based and optimization methods which require no training. Filter-based methods focus on preserving sharp depth edges under the guidance of the intensity image. For example, RGB images guide the acquisition of bilateral weights \citep{kopf2007joint}. Optimization-based approaches use diverse data priors to construct energy functions, with data-fidelity regularization limiting the solution space. Several other methods \citep{xie2015edge,diebel2005application} employ random field models. Other than that \citet{liu2013guided} demonstrated an early learning-free application of anisotropic diffusion. Initial learning-based techniques, such as bimodal co-sparse analysis \citep{kiechle2013joint} and joint dictionary learning \citep{tosic2014learning}, learn the relationship of RGB and depth information. Deep learning models are introduced to utilize neural networks to learn the mapping from low-resolution to high-resolution images. MSG-Net \citep{hui2016depth} built on a U-Net deep network architecture and learns the residual errors of bicubic interpolations by embedding the source at the smallest scale. \Citet{kim2021deformable} propose \gls{gl:DKN} and its fast implementation (FDKN), that learn sparse and spatially-invariant filter kernels. \Citet{He2021} employ a high-frequency guidance module to embed the guide details into the depth map. Deep learning techniques have been used recently by researchers to improve the ability of predicting outputs of given inputs that have not encountered within formal frameworks. \Citet{riegler2016deep} train a neural network by unrolling the optimization phases of a first-order primal-dual algorithm, enabling them to train their deep feature extractor throughout the training process. Using a graph-based, MRF-style optimizer, \Citet{Lutio2022} apply the implicit function theorem. DADA \citep{Metzger2023} achieved state-of-the-art performance by adapting the concept of guided anisotropic diffusion with deep convolutional networks.

\subsection{DSM Super-Resolution}

Research on \Gls{gl:DSM} super-resolution is limited, despite its significance in remote sensing. The majority of research focuses on DEM superresolution. This includes interpolation-based methods like bicubic, and bilinear are used for DEM enhancement which results in smooth terrain models. \Citet{xu2015nonlocal} proposed a super-resolution algorithm based upon non-local means. It operates by using a predetermined equation to search for similar patches across the training set. Weights determined by the searching phase are then used to upscale the resolution of the target \gls{gl:DEM}. Deep learning-based methods also derived from the advances in \gls{gl:SISR} domain. D-SRCNN \citep{chen2016convolutional}  was proposed based on SRCNN \citep{dong2015image}. Later the same author implemented EDSR \citep{lim2017enhanced} for the same purpose \citep{xu2019deep}. Demiray et al. proposed a \gls{gl:DEM} super-resolution model, namely D-SRGAN \citep{demiray2021d}, with  the implementation of SRGAN~\citep{ledig2017photo}, and EffecientNetV2 \citep{tan2021efficientnetv2}  for \gls{gl:DEM} SR \citep{demiray2021super}. 

So far none of the methods targeted specifically for urban \glspl{gl:DSM}. Because of their characteristics, performing large-scale super-resolution on such data is still a challenge. First, these mentioned models sofar trained only on the data generated by bicubic interpolation which only work well on clean low-resolution data with simple degradations. This is inconsistent with real-world needs, where low-resolution data have more complex degradations. Second, urban DSMs provide more details which even harder to reconstruct. To address this conflict, we proposed a practical solution, where we collect real low and high-resolution DSMs pairs and include guiding information in both training and inference. Optical images of the same scenes are obtained to act as guidance. Other than that, We utilize both local and global context in our approach to handle the characteristics of DSMs. In the following, we explain in more detail how we achieve this. 

\section{Methodology}

\RealGDSR consists mainly of two steps: First as a pre-step, a pre-trained model is used to extract features from both optical image and low-resolution \gls{gl:DSM}. Then, we use a residual \gls{gl:CNN}-based network, namely the local refinement network, whose objective is to take as input high-dimensional features and transform them into a high-resolution \glspl{gl:DSM}. Second, we use an anisotropic diffusion network to further enhance the refined \glspl{gl:DSM} focusing on the edge features from the high-resolution optical image. \Cref{fig:realgdsr} shows an overview of our proposed pipeline. In the following subsections, we describe each component of our pipeline in more details.
\subsection{Local Refinement}
In prior works in DEM super-resolution, we observe that many existing works often follow a common design concept for image super-resolution, where their networks built of upsampling layers and have very large receptive field, for example, a U-Net like architecture or using multiple dilation convolution or attention layers. In this work, taking inspiration from image inpainting task, we see super-resolution as an image-to-image translation task where we assume that an initial coarse high-resolution \gls{gl:DSM} of the observed scene has already been generated with existing traditional method like bicubic and later refined within the training. We highlight that for \glspl{gl:DSM} a network with small receptive field is enough to fulfill the task. For the local refinement, we create a shallow network with four residual blocks and two upsampling procedures (see middle part of \cref{fig:realgdsr}). Therefore, this network has small receptive fields. This architecture eliminates the effect of distant and unsuccessful filling contents and allows for the restoration of missing sections or buildings utilizing local information around them. Additionally, we incorporate a long skip connection that adds the initial \gls{gl:DSM} straight to the final downsampling layer's output, allowing the network to regress residuals rather than absolute height.


\subsection{Edge-Enhancing Diffusion}\label{sec:Footnotes}

After the local refinement process, structural features are restored with guidance of surrounding local regions. However, different than images, \glspl{gl:DSM} are sensitive to outliers. For this purpose, we introduce a diffusion-based edge-enhancement which helps broaden the scope of information captured. Anisotropic diffusion can be understood as an adaptive filtering technique aimed at smoothing while preserving inter-region content such as edges or boundaries, which are crucial for image interpretation. This is achieved by applying inhomogeneous diffusion, where the diffusivity is guided by a scalar function or diffusion tensor derived from the gradients of the evolving image. The concept of calculating diffusivity from a guide image has been investigated in the area of edge enhancement and semantic segmentation.

Inspired by prior work in guided depth super-resolution \citep{Metzger2023}, we implement a similar network, where the diffusion component mirrors traditional optimization approaches solved via an iterative diffusion loop. For each iteration, multiple steps of anisotropic diffusion are conducted. Here, the diffusion weights are influenced by the guide to minimize diffusion at boundaries with high contrast and enhance diffusion within regions that are homogeneous. A convolutional feature extractor is used to set diffusion weights by passing the guide through. Therefore, the process can transfer edge information from the guide to preserve depth discontinuities in the target image.

Given a source image $S$ a guide $G \in H \times W \times C$, where $C = 3$ for RGB images or a larger
number for deep features, the first step is to initialize $X_0 \in H \times W$ with an upsampled version of $S$. The diffusion step is defined as
\begin{equation}
\hat{x}^p_t = x^p_{t-1} + \lambda \cdot \sum_{n \in N_4(p)} (x_{t-1}^n - x_{t-1}^p)   \cdot c(g^p, g^n) ,
\end{equation}
where $x^p_t$ denotes the pixel value of $X_t$ at location $p$ (and similarly for $g^p$). $N_{4(p)}$  denotes the four-neighbours of pixel $p$. $\lambda$ controls the rate of diffusion. For four-neighbours, $\lambda$ should be set to  $< 0.25$ to ensure numerical stability. Diffusion coefficients for the neighboring pairs of pixels are calculated by function $c$ which formulated based on their values in the guide. A higher diffusion coefficient means that information spreads more freely across neighboring pixels, resulting in stronger smoothing effects. Conversely, a lower diffusion coefficient restricts the spread of information, preserving edges and fine details in the image.
This method follows prior work \citep{perona1990scale}, which defines

\begin{equation}
c(g^p, g^n) = \frac {K^2} {K^2 + \mid\mid {g^p - g^n} \mid\mid ^2_2} 
\end{equation}

where $K$ controls the sensitivity to the gradients in $G$.

We adapt the work of \citet{Metzger2023} by removing the adjustment step. The adjustment step was utilized by the authors to constrain the output of the diffusion to always match the source image when downsampled. This is done to preserve adherence between the input and output. However, in our approach, we opt to forego the adjustment step. This decision is motivated by the recognition that low-resolution DSMs often contain minimal information, and their use in the adjustment step may hinder the diffusion process. 

To this end, our proposed network is trained in an end-to-end manner, and the final training loss is the summation of losses of two sub-networks.

\section{Experiments}
\newcommand{\dummyfigure}{\tikz \fill [NavyBlue] (0,0) rectangle node [black] {Figure} (2,2);}
\newcolumntype{M}[1]{>{\centering\arraybackslash}m{#1}}
\begin{figure*}[t]
    \centering
    \begin{tabular}{cM{50mm}M{50mm}M{50mm}}
        \toprule
        Model & Area 1 & Area 2 & Area 3  \\
        \midrule
        \rotatebox[origin=c]{90}{Bicubic} 
        & \includegraphics[width=\linewidth]{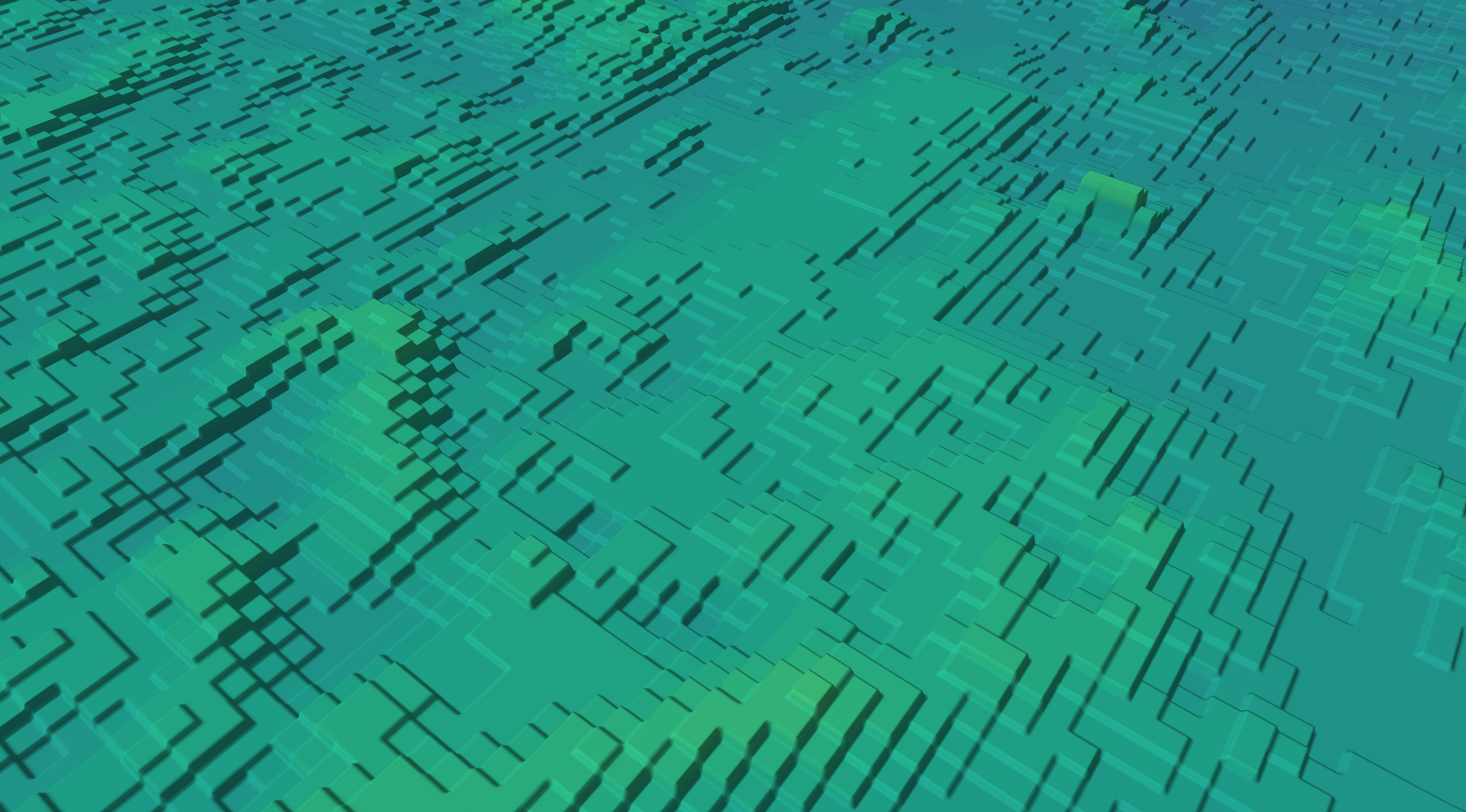} 
        & \includegraphics[width=\linewidth]{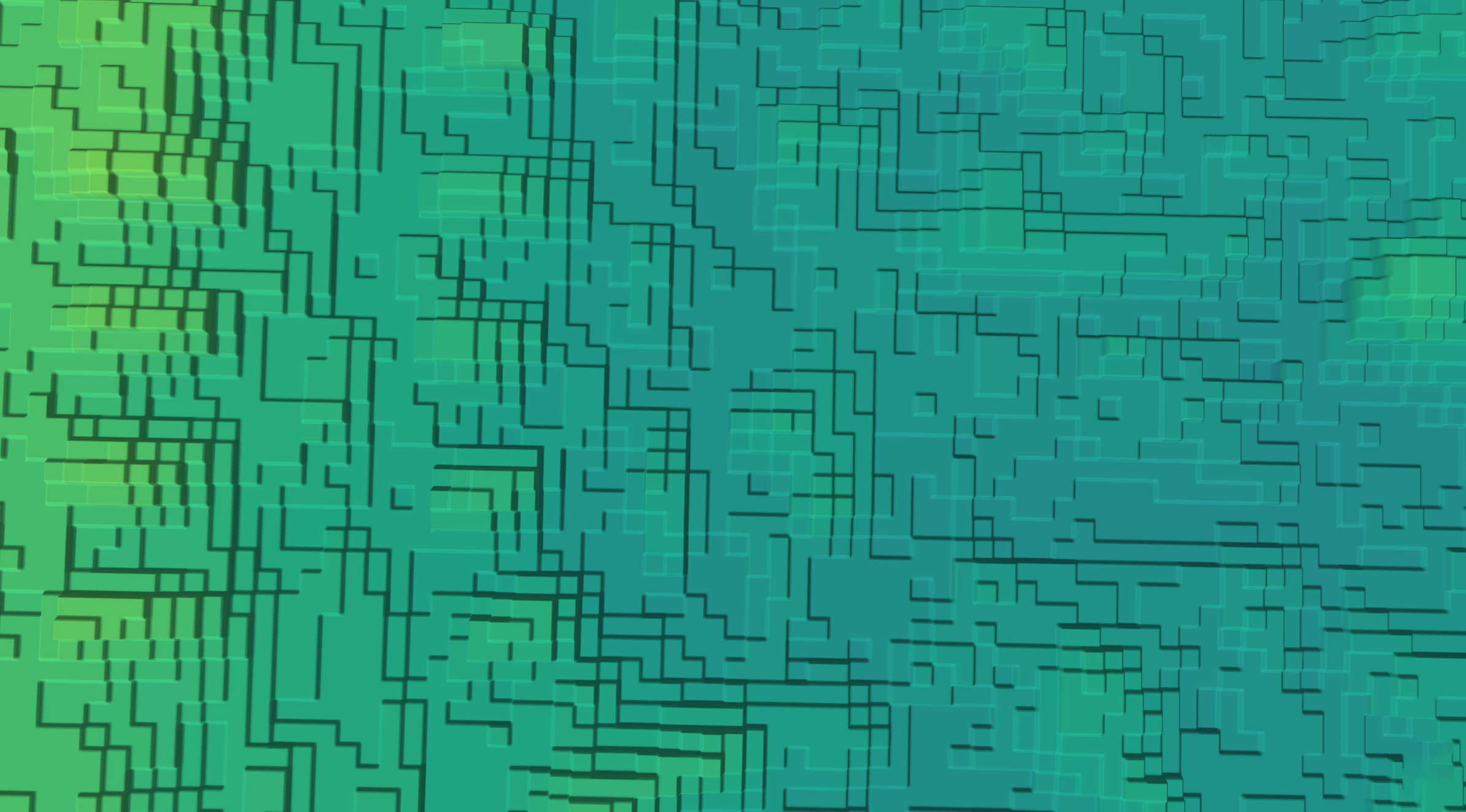} 
        & \includegraphics[width=\linewidth]{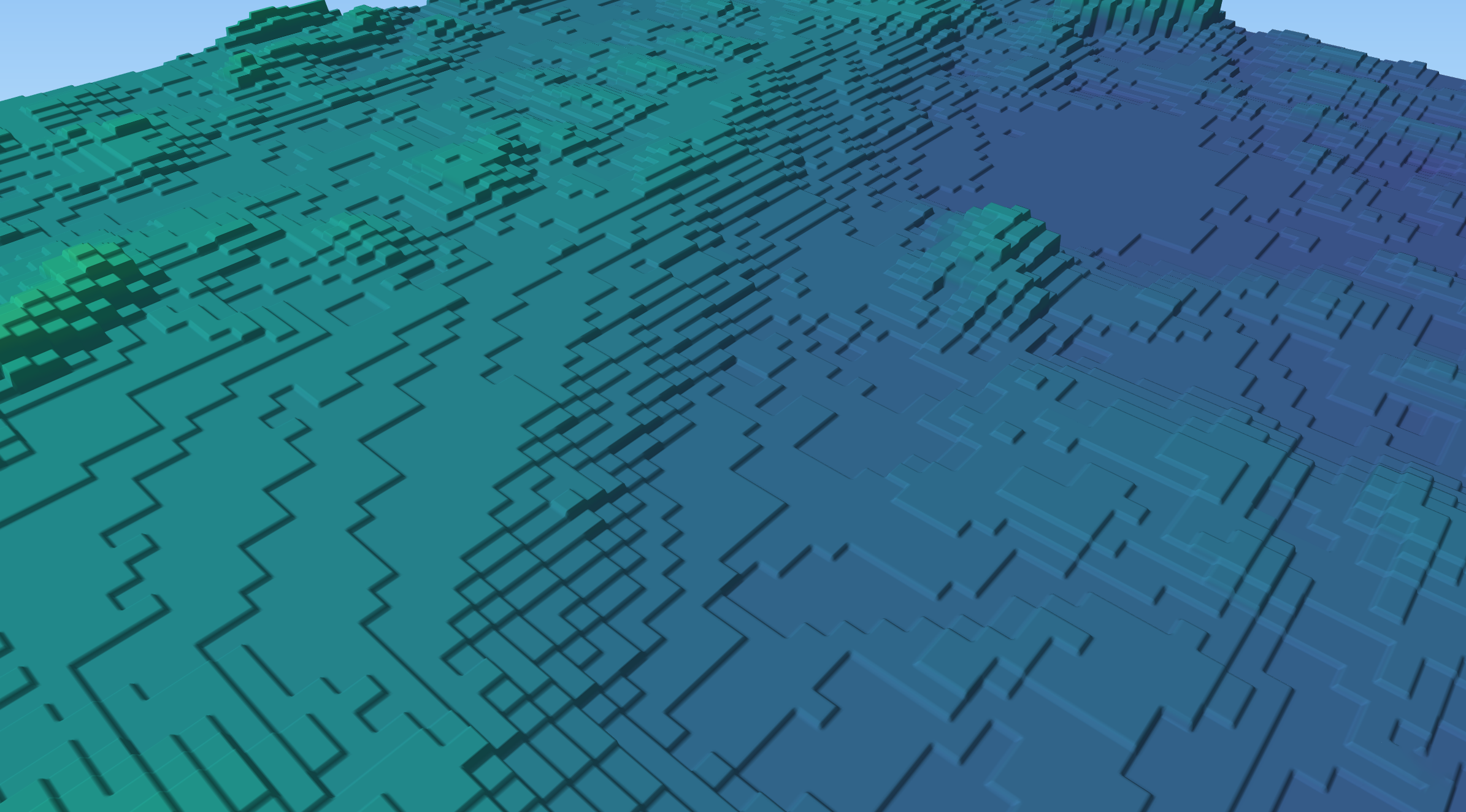} \\
        \rotatebox[origin=c]{90}{DADA} 
        & \includegraphics[width=\linewidth]{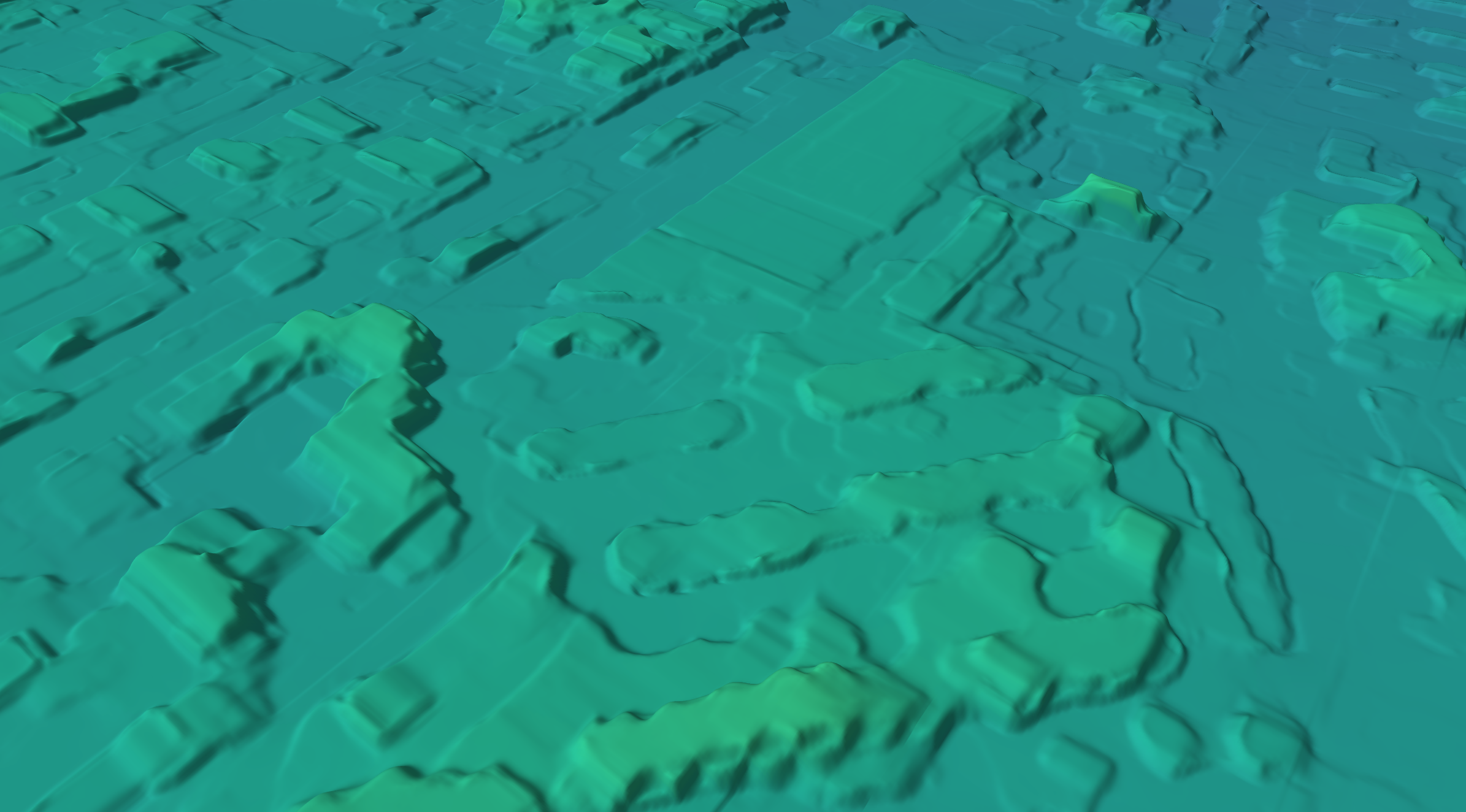} 
        & \includegraphics[width=\linewidth]{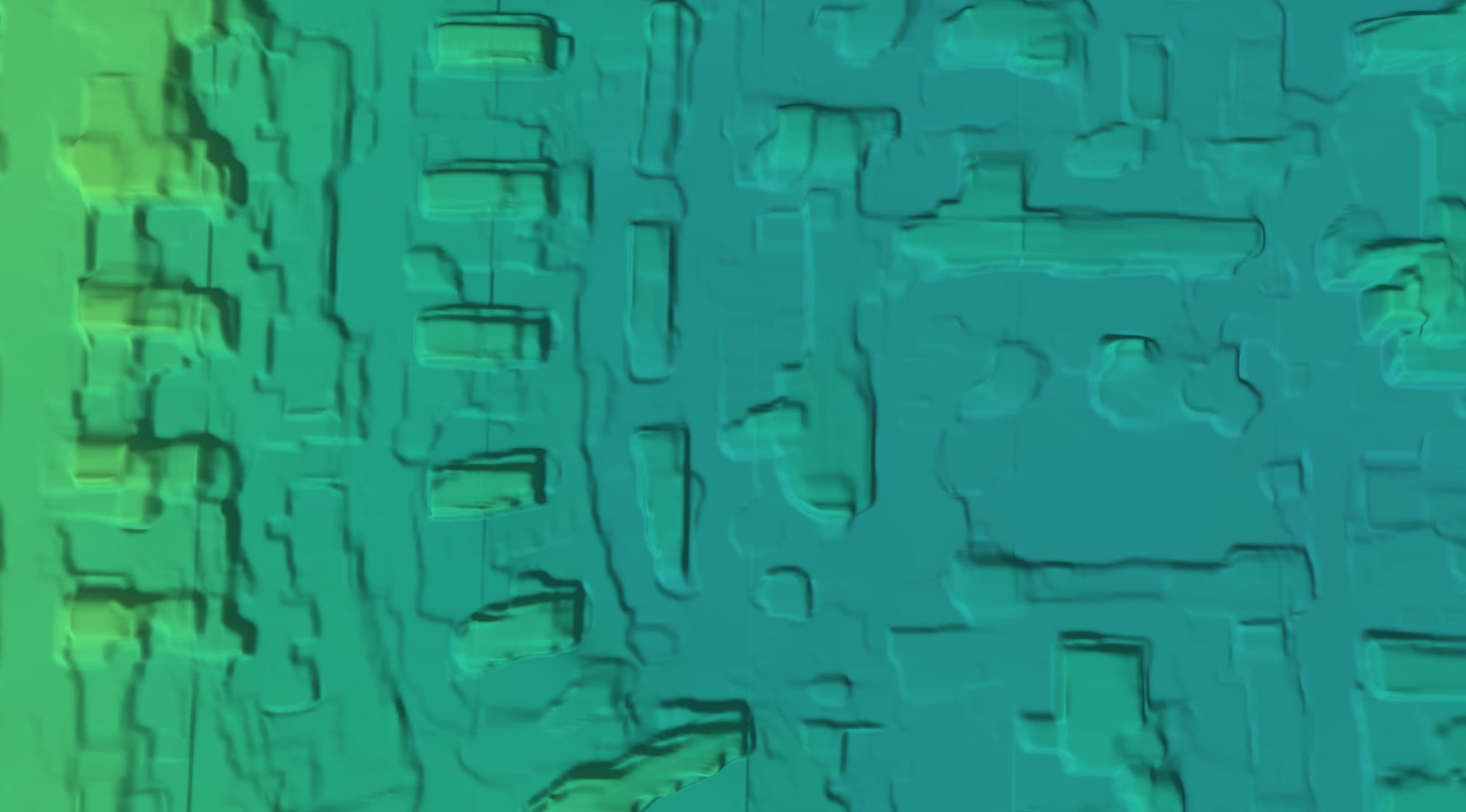} 
        & \includegraphics[width=\linewidth]{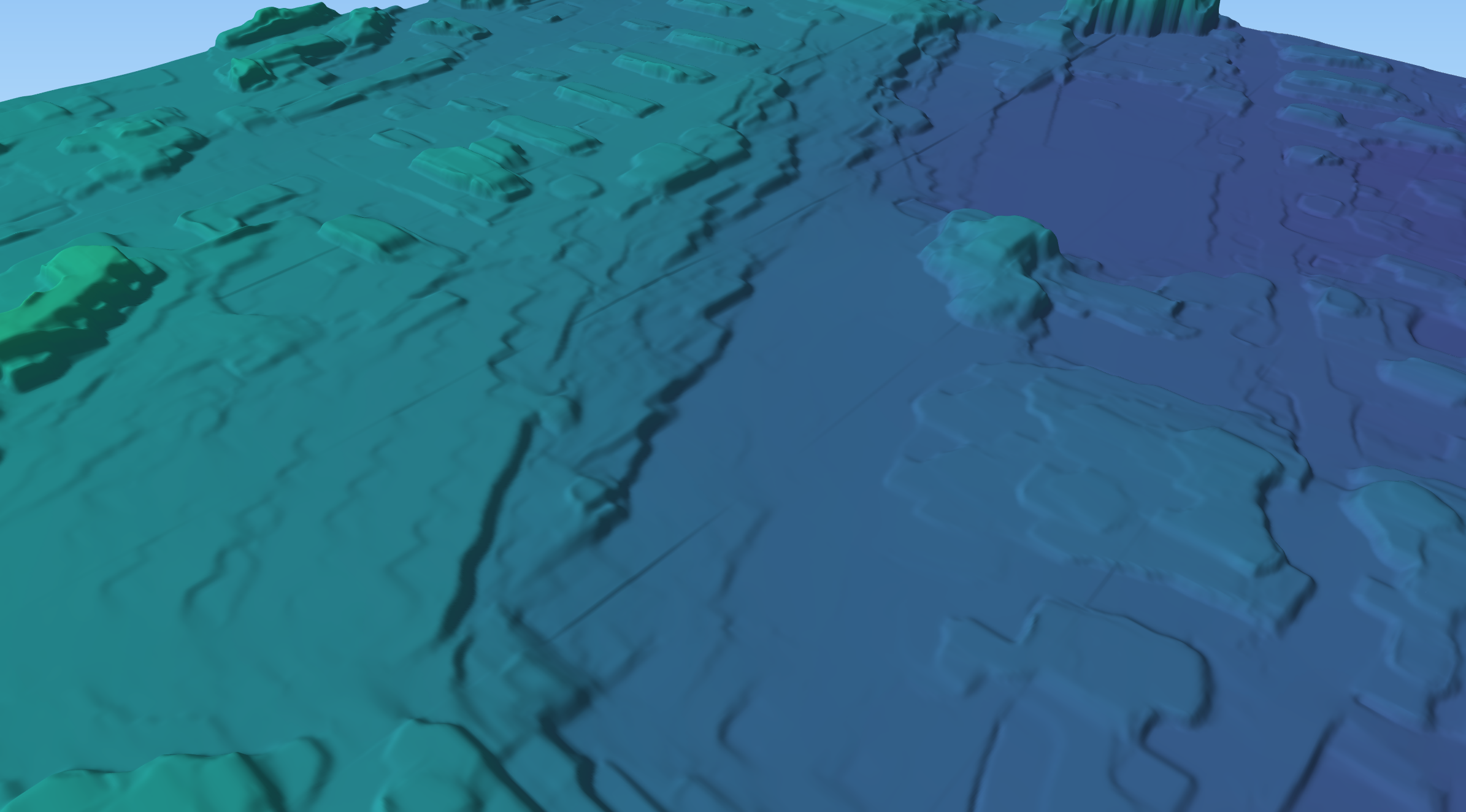} \\
        \rotatebox[origin=c]{90}{DSRGAN} 
        & \includegraphics[width=\linewidth]{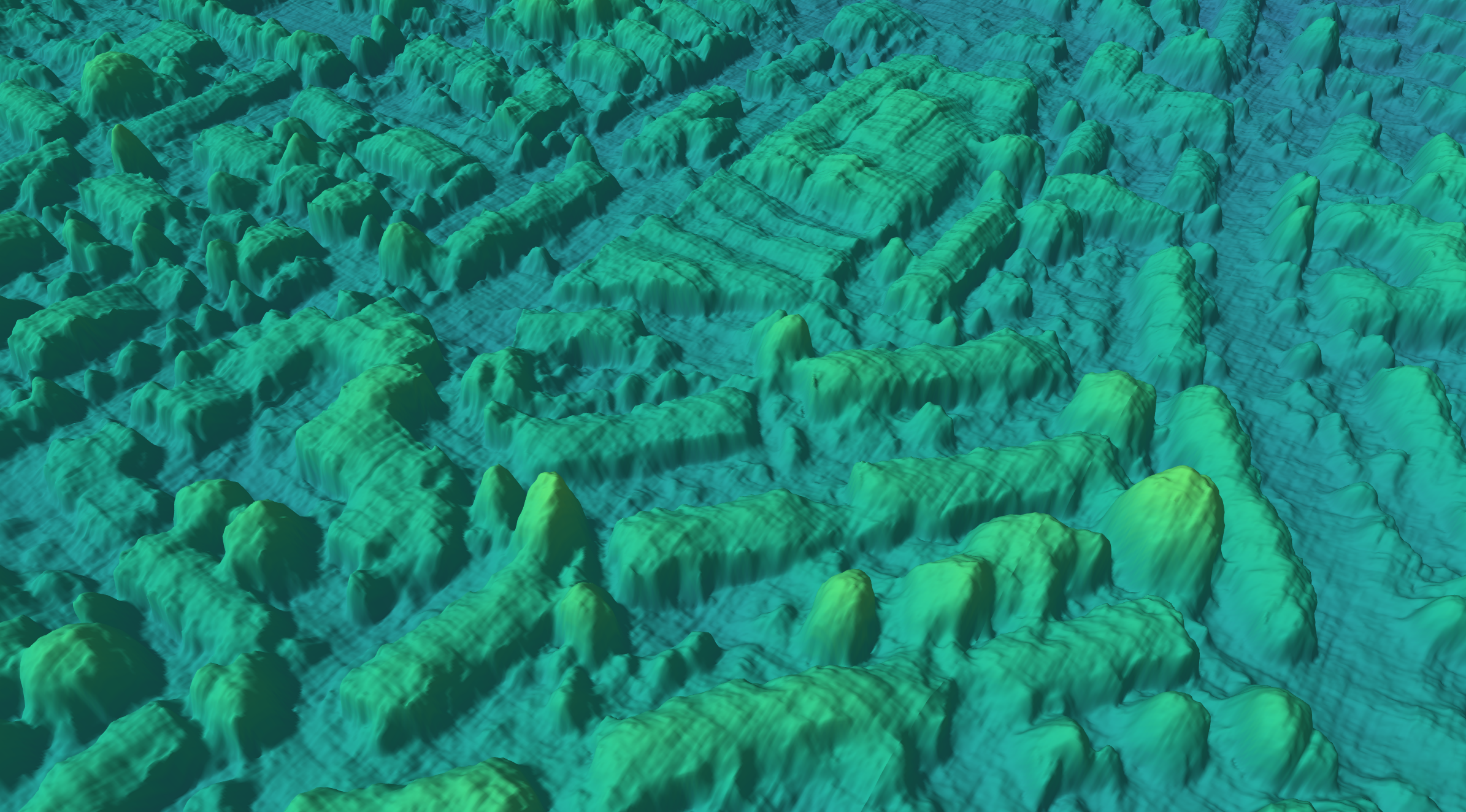} 
        & \includegraphics[width=\linewidth]{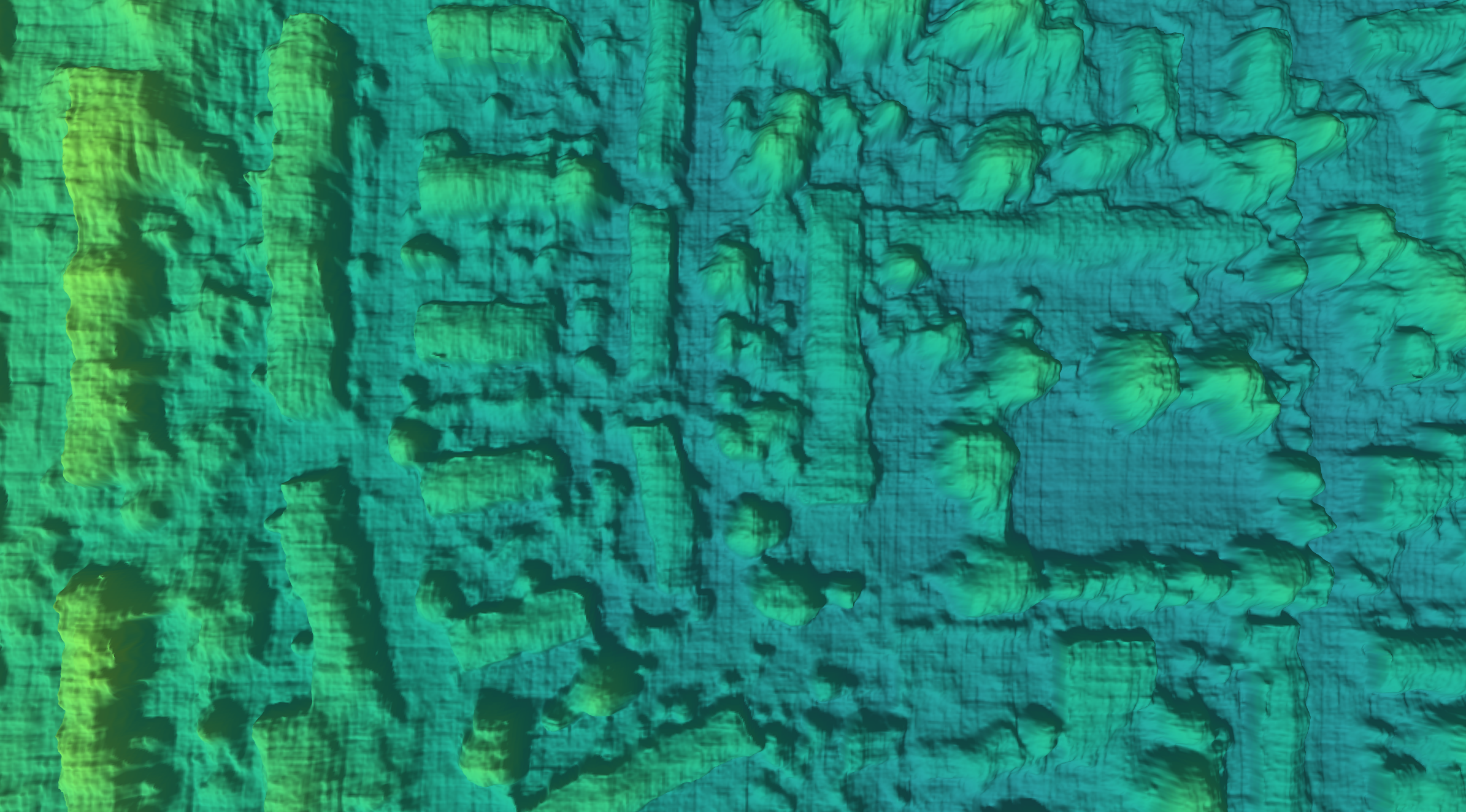} 
        & \includegraphics[width=\linewidth]{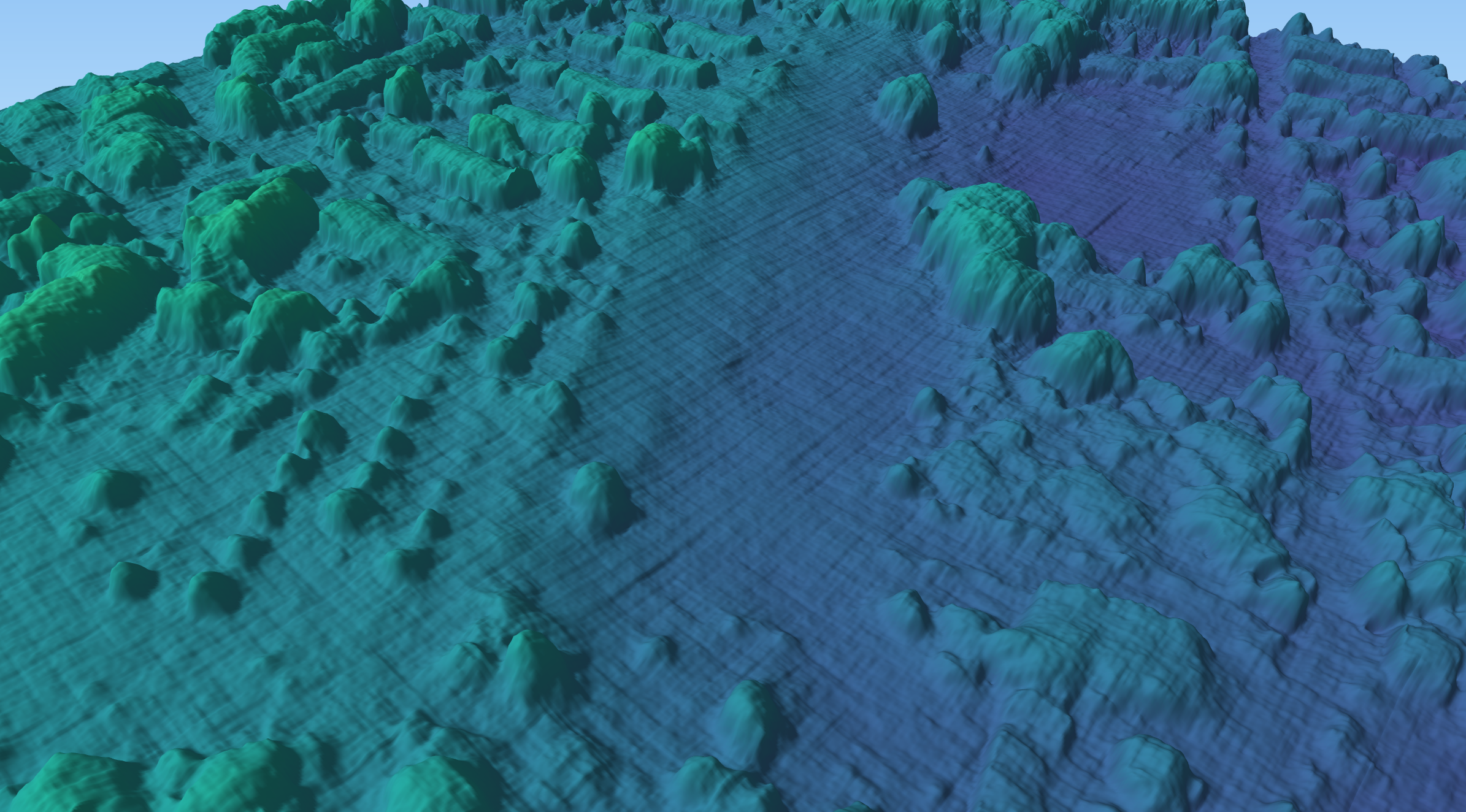} \\
        \rotatebox[origin=c]{90}{RealGDSR} 
        & \includegraphics[width=\linewidth]{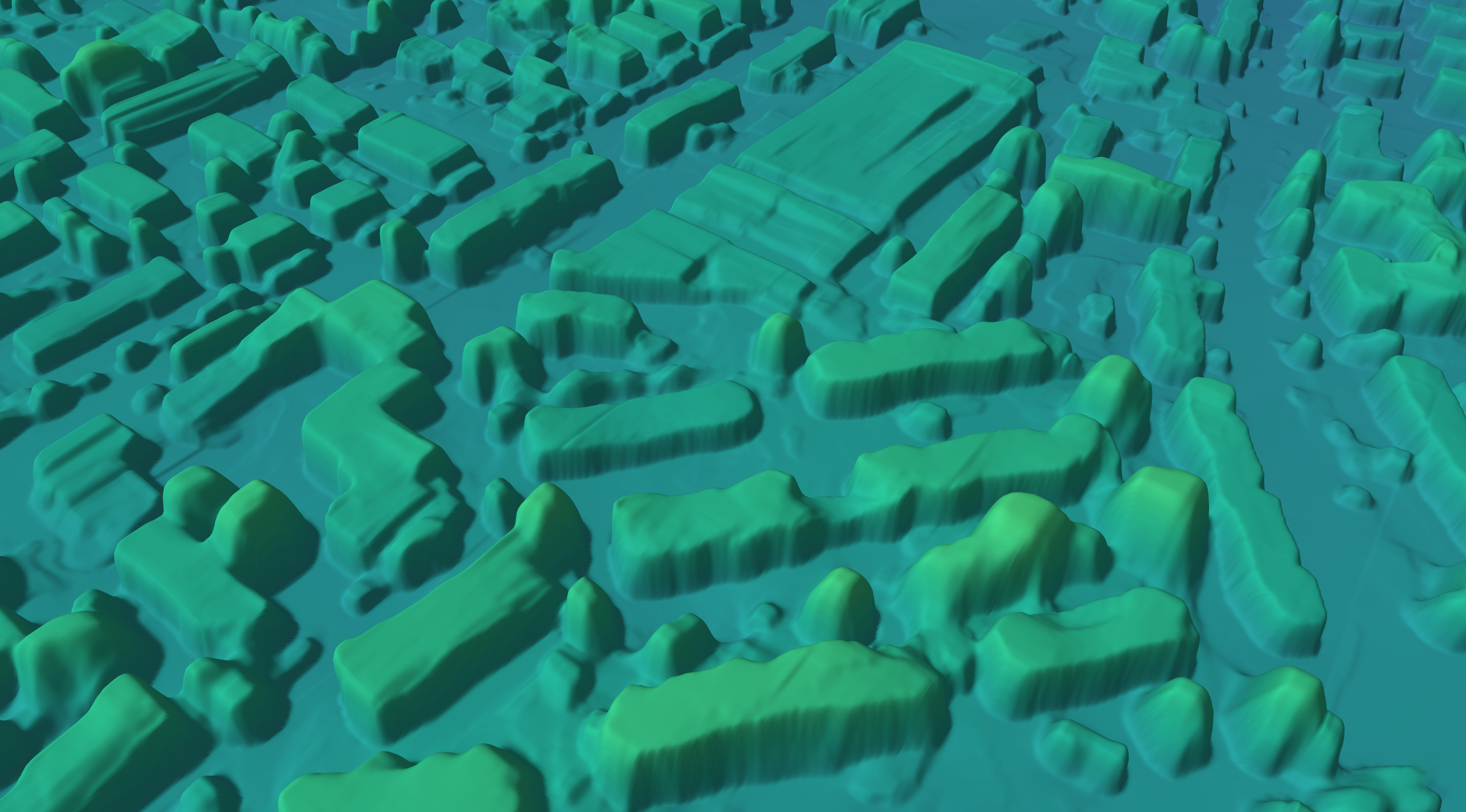} 
        & \includegraphics[width=\linewidth]{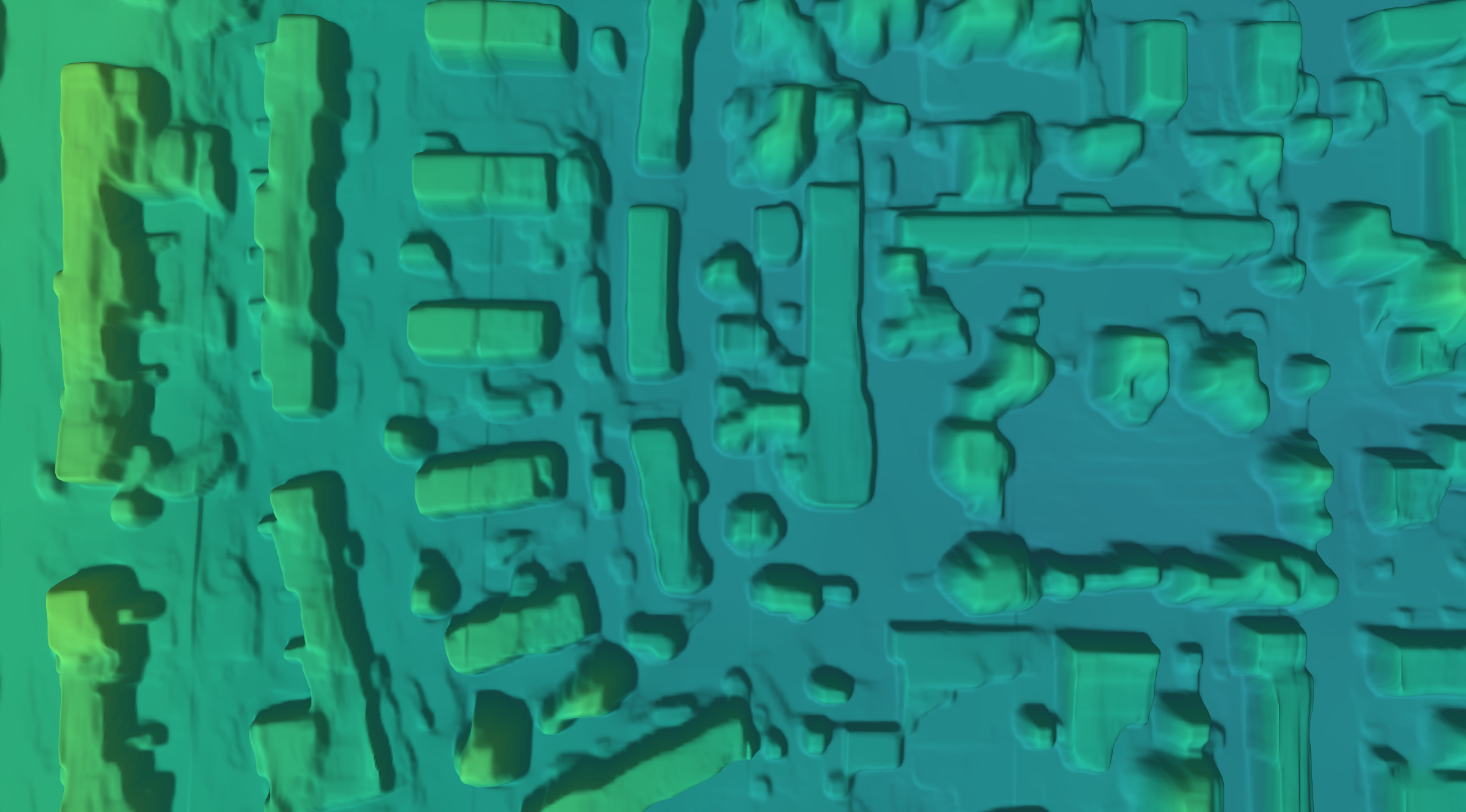} 
        & \includegraphics[width=\linewidth]{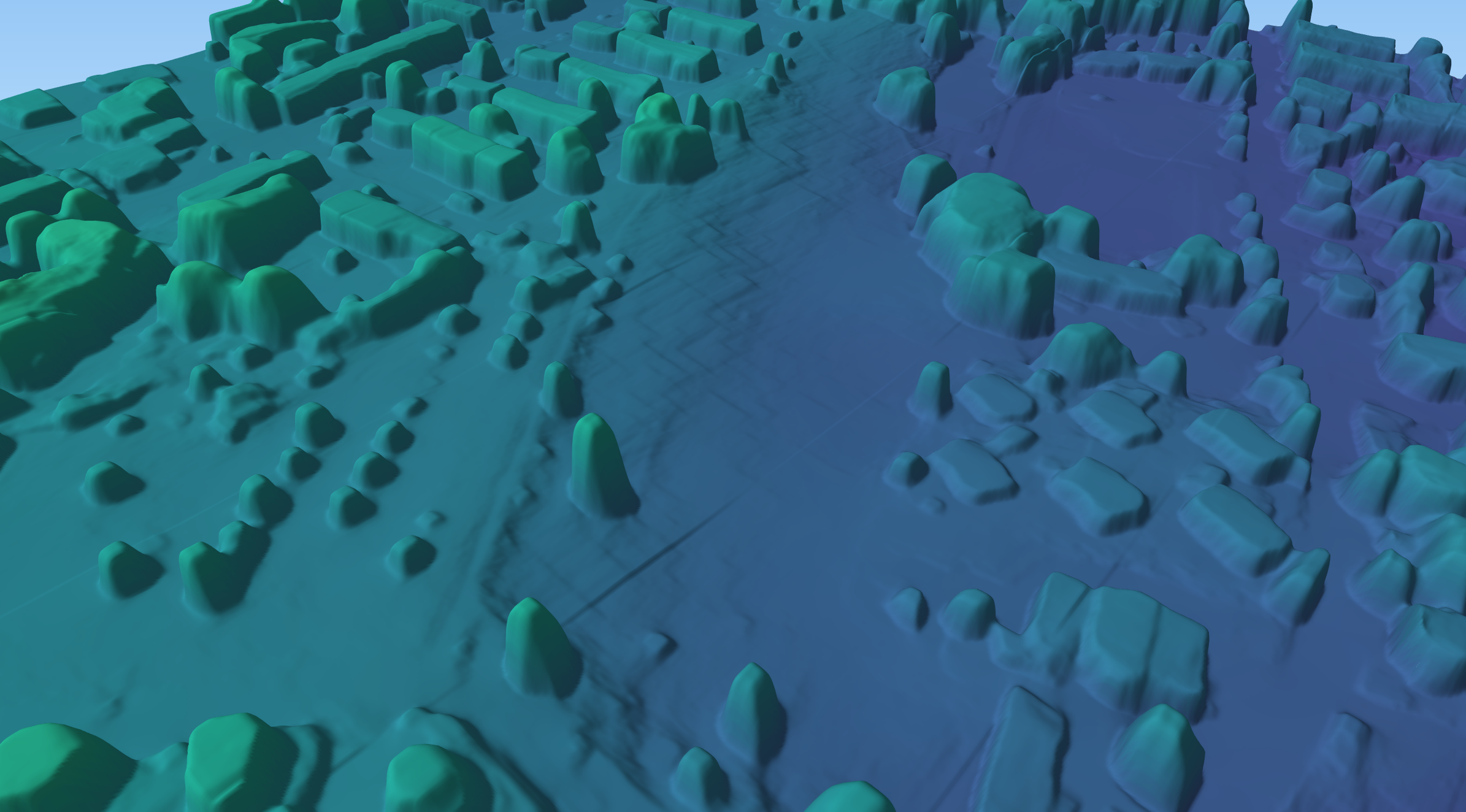} \\
        \rotatebox[origin=c]{90}{GT} 
        & \includegraphics[width=\linewidth]{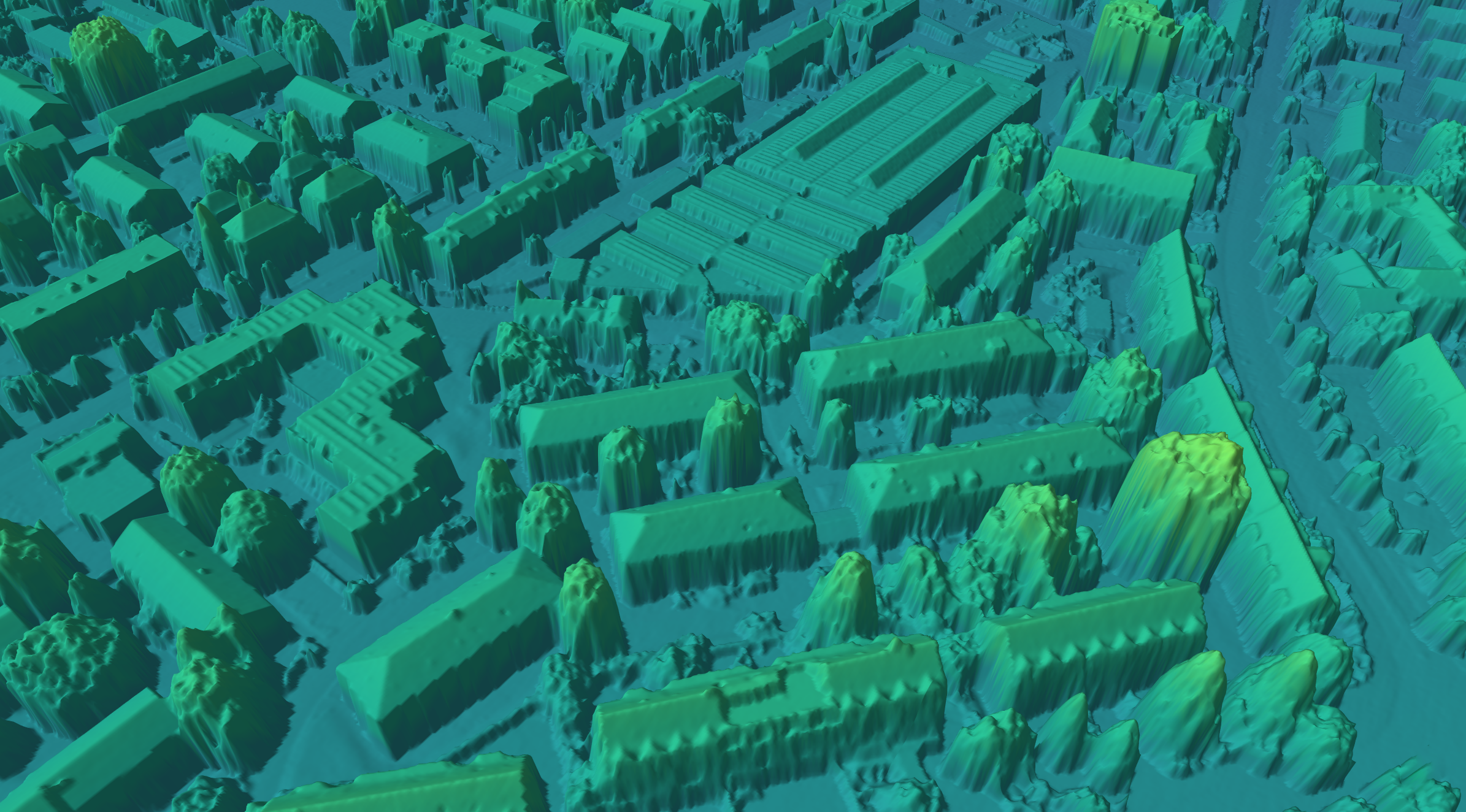} 
        & \includegraphics[width=\linewidth]{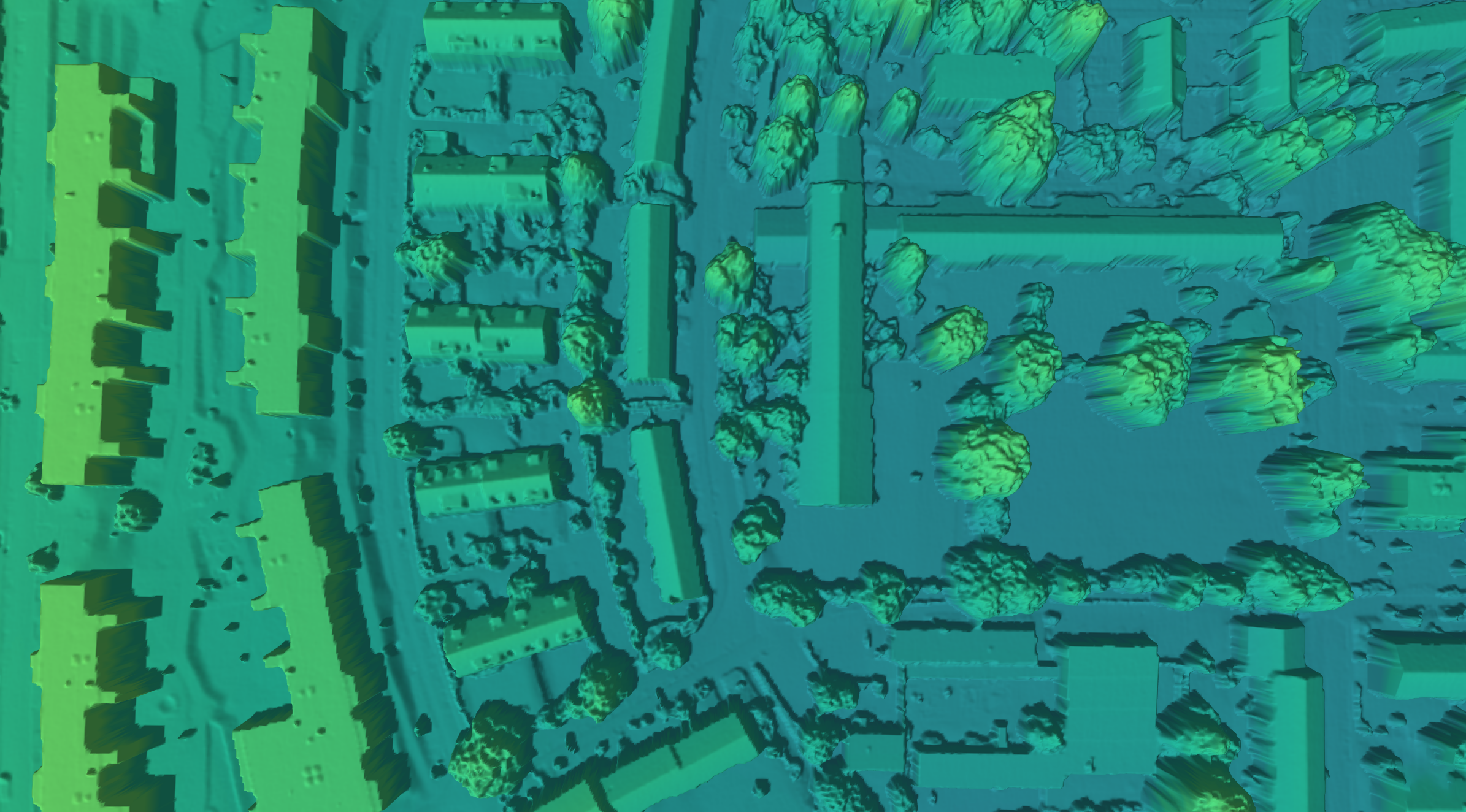} 
        & \includegraphics[width=\linewidth]{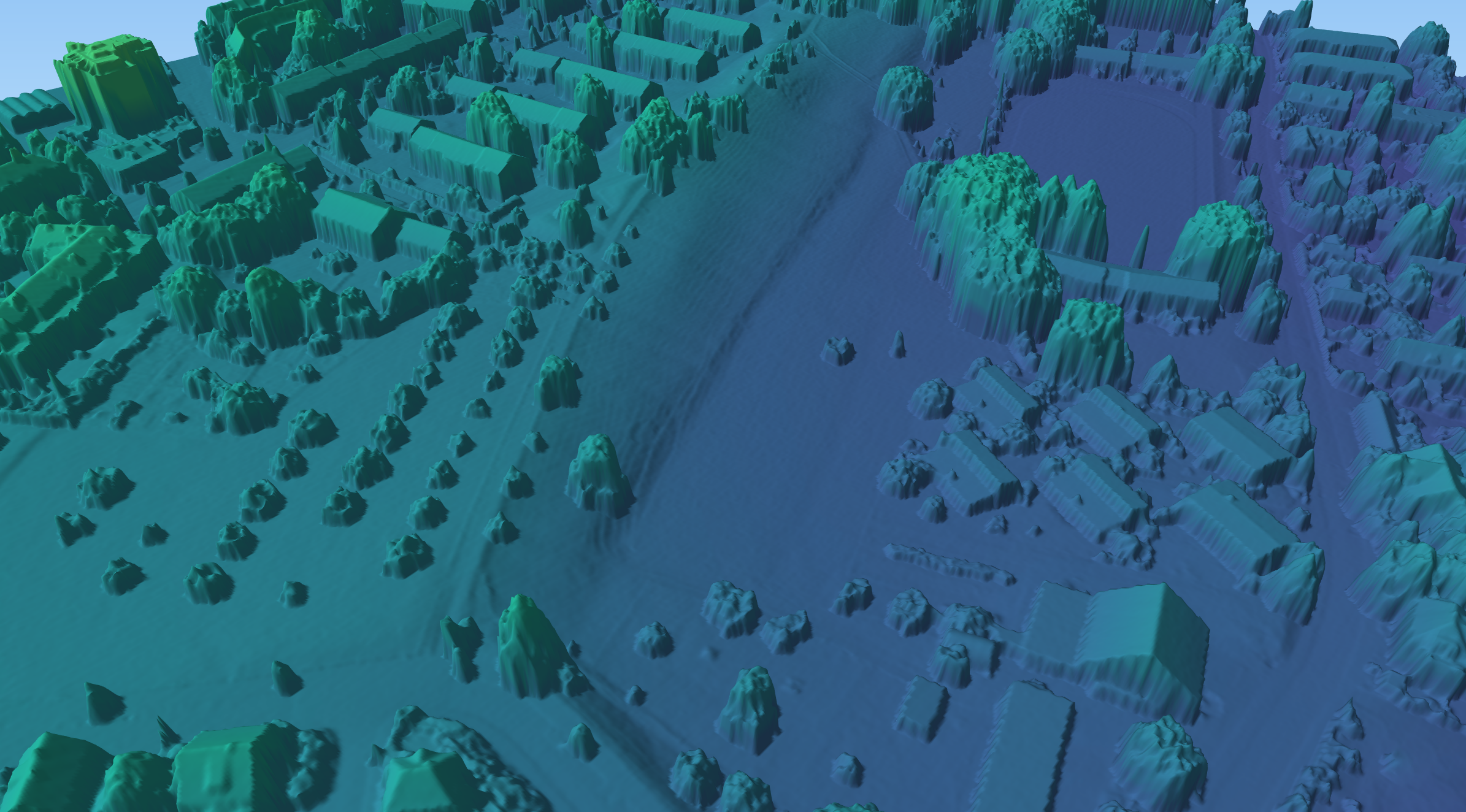} \\

        \bottomrule
    \end{tabular}
    \centering
    \caption{Visual comparison of RealGDSR with selected baselines. Our approach demonstrates its accuracy while producing regularized and smooth DSMs. All examples are taken from the test set.}
    \label{fig:comparison-3d}
\end{figure*}

\subsection{Datasets and Implementation}
\paragraph{Imagery and Study Area} In this study, we generate a dataset based on real-world DSMs. We evaluate our method on \glspl{gl:DSM} acquired over two main cantons of Switzerland: Zurich and Bern. We use low-resolution \gls{gl:DSM} with a \gls{gl:GSD} of \SI{5}{\meter} generated from Cartosat-1 stereoscopic satellite as input instead of conventional bicubic-downsampled. We take high-resolution \gls{gl:DSM} and their corresponding RGB images (\gls{gl:GSD}=\SI{0.5}{\meter}) provided by The Federal Office of Topography on the Swisstopo Portal\footnote{\url{https://www.swisstopo.admin.ch/en/geodata.html}}. The dataset consists of \SI{2200}{} patches of size (\SI{256}{}, \SI{256}{}). We use \SI{2000}{} samples for training and \SI{200}{} for testing, The study area includes widely spaced, detached residential buildings, allotments, and high commercial buildings. We accomplish this by utilizing the Esri global land cover map \citep{karra2021global} to filter the dataset, specifically focusing on the \textit{Built Area} class.

\paragraph{Implementation Details}
 We randomly load training patches during training. At inference time, we reconstruct large-scale scenes by applying the learned model in a sliding window. Every patch is scaled such that all height points averaged to \SI{0}{} using its local mean and global standard deviation computed from all training samples. Optical images are normalized with the mean and standard deviation over the intensity values of all training pixels. In all experiments, we use a hidden feature dimension of \SI{64}{} for the feature extractor and the refinement decoder. ResNet-50 \citep{he2016deep} backbone pretrained on ImageNet \citep{deng2009imagenet} is used as feature extractor. Before extracting the features, The low-resolution DSMs are upsampled to the resolution of the optical images using bicubic interpolation and concatenated with the guide. We follow best practice and normalize the data for neural network training For training, we train all methods, including our own, with the $\mathcal{L}_1$ loss. For the diffusion network, we adopt the same setup and strategy outlined in \citep{Metzger2023}, with $K$ and $\lambda$ set to $0.001$ and $0.24$. The number of diffusion steps with and without gradients in training phase, are set to $8000$ and $1024$, respectively. Additionaly in our refinement network and DSRGAN, perceptual loss is added. We employ the ADAM optimizer with a base learning rate of $5 \times 10^{-5}$ and no weight decay. We set the batch size to $2$ for training and $1$ for testing. We stop training once the RMSE on the test set have converged. We implemented our model in PyTorch and run it on a NVIDIA Titan RTX GPU. 
 
\medskip

\subsection{Baselines}

We compare \RealGDSR against the following baselines:
\begin{itemize}
\item Bicubic: The upsampled low-resolution \glspl{gl:DSM}  using bicubic-interpolation.
\item DADA: Deep Anisotropic Diffusion-Adjustment
network is a hybrid framework for guided super-resolution that combines deep feature learning and anisotropic diffusion. The approach achieved edge-enhancing properties from the diffusion boosted by the contextual reasoning capabilities of large pre-trained models and a strict adjustment step guarantees perfect adherence to the source image. Our diffusion network is similar to this approach without the adjustment step.

\item D-SRGAN: DEM Super-Resolution with Generative Adversarial Networks implemented ESRGAN for single \gls{gl:DSM} super-resolution. To compare it with our models we modify the input of the model by concatenating the low-resolution \gls{gl:DSM} with extracted features from the optical images using the same feature extractor as our model. We make no changes to the GAN.
\end{itemize}

\subsection{Evaluation Metrics} 
\label{sec:metrics}
We evaluate the models' performance by examining the root mean square error (RMSE), the normalized median absolute deviation (NMAD), and the median absolute error (MedAE), which are all derived from per-pixel differences between predicted and ground truth.

\section{Results}
\subsection{ Comparisons with Prior Works}
\begin{figure}[!t]
    \centering
    
    \begin{subfigure}{.28\linewidth}
    \centering
    \includegraphics[width=\linewidth]{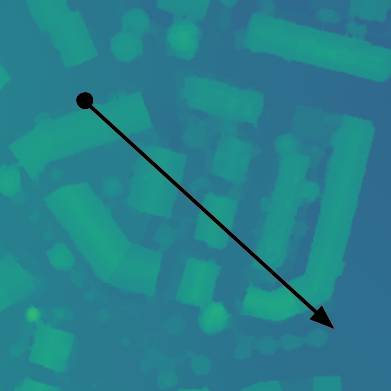}
    \vspace{10pt}
    
    \caption{Area of interest}
    \label{fig:fig:driving2kitti}
    \end{subfigure}
    \hfill
    \begin{subfigure}{.38\linewidth}
    \centering
        \begin{tikzpicture}[scale=0.50]
            \begin{axis}[
                xlabel={$Distance$},
                ylabel={$Height$},
                grid=major,
                legend pos=outer north east,
                legend style={align=left},
                grid=none,
                xmin=0,   
                xmax=110, 
                ymin=444.78, 
                ymax=464.14, 
            ]
    
            \addplot [darkgray, no markers]  table {plots/GT.txt};
            \addplot [blue, no markers, dashed]  table {plots/Bicubic.txt};
            \addplot [green, no markers, dashed]  table {plots/DADA.txt};
            \addplot [orange, no markers, dashed]  table {plots/DSRGAN.txt};
            \addplot [red, no markers, dashed]  table {plots/REALGDSR.txt};
    
            \addlegendentry{GT}
            \addlegendentry{Bicubic}
            \addlegendentry{DADA}
            \addlegendentry{DSRGAN}
            \addlegendentry{REALGDSR}
    
            \end{axis}
        \end{tikzpicture}
    \captionsetup{margin={1.25cm,0cm}}
    \caption[caption]{Line profile}
    \label{fig:fig:synt2real}
    \end{subfigure}
    \hspace{65pt}
    \caption{Line profile analysis of RealGDSR and other baselines.}
    \label{fig:line-profile}
\end{figure}

We start by assessing the performance of \RealGDSRdot, so as to quantify the impact of our framework. The high-resolution \gls{gl:DSM} generated by bicubic interpolation serves as the baseline. Due to the limited transformation, the bicubic-upsampled \glspl{gl:DSM} are blocky and lack of structural features (see \cref{fig:comparison-3d}, 1st row). The RMSE is \SI{5.6}{\meter} and the MedAE is \SI{2.3}{\meter}. Applying our model improves the reconstruction significantly. The RMSE is lowered to \SI{3.6}{\meter}, a 50\% improvement compared to the bicubic-upsampled. Similarly, the NMAD and MedAE are also reduced to \SI{2.6}{\meter} and  \SI{1.5}{\meter}.
Beside the quantitative improvement, visually we can see that the reconstructed 3D geometry is clearly recovered. Buildings have sharp lines, and there are fewer visible artifacts and bumps on the terrain. The most notable finding is the recovery of detailed building structures like a cluster of buildings on \cref{fig:comparison-3d}, 4th row, 1st column. 
Furthermore, \RealGDSR reconstructs realistic scenes even in the presence of previously unseen building shapes and arrangement of buildings. The reconstruction of height and arrangement of buildings in urban areas are more accurate, where most of the baselines failed (see \cref{fig:line-profile}). In such cases, the model predict the height using information extracted from the optical image. 

\begin{table}[!ht]
    \centering
    \begin{tabular}{lccccc}
        \toprule
        Model & RMSE & NMAD & MedAE\\
        \midrule
        Bicubic  & 5.59 & 3.62 & 2.30 \\
        \midrule
        DADA & 5.55 & 3.40 & 2.17 \\
        D-SRGAN & 4.1 & \textbf{2.06} & 3.08 \\
        \midrule
          Real-GDSR & \textbf{3.54} & 2.65 & \textbf{1.46} \\
        \bottomrule
    \end{tabular}
    \caption{Performance comparison of \RealGDSR and baselines in meters (m). Our model outperforms other methods.}
    \label{tab:comparison-metrics}
\end{table}

Among all methods \RealGDSR yields the lowest reconstruction errors (\cref{tab:comparison-metrics}). Compared to other networks it reconstructs smoother surfaces and more accurate building heights and features (\cref{fig:comparison-3d}, 4th row) where other networks failed to either predict accurate height (\cref{fig:comparison-3d}, 2nd row) or produce noises on the terrain and building shapes (\cref{fig:comparison-3d}, 3rd row). Quantitatively, the difference amounts to 15\% in RMSE in comparison to the second best model, DSRGAN.

\subsection{Ablation Study}
\begin{table}[!ht]
    \begin{center}
    \begin{tabular}{l|cc|ccc}
        \toprule
        & Refinement & Diffusion & RMSE & NMAD & MedAE \\
        \midrule
        1 & $\checkmark$  & & 4.54 & \textbf{1.81} & \textbf{1.42}\\
        2 &  & $\checkmark$ & 5.53 & 3.45 & 1.95\\
        \midrule
        3 & $\checkmark$  & $\checkmark$ & \textbf{3.54} & 2.65 & 1.46\\
        \bottomrule
    \end{tabular}
    \caption{Contribution of each component in our network}
    \label{tab:ablation}
    \end{center}
\end{table}

\paragraph{Local Residual Refinement Network} 
One of the main contributions of this work is our local residual refinement step of bicubic-interpolated sample which focusing on local features. To evaluate this we retrained our model removing the diffusion step leaving only refinement network and compared it (\cref{tab:ablation}, 1st row) with the modified DSRGAN (\cref{tab:comparison-metrics}, 3rd row). Although having worse RMSE with \SI{0.4}{\meter}  difference, visually our network is able to produce more regularized \glspl{gl:DSM} with less visible artifacts, which is shown by advantage over DSRGAN in NMAD and MedAE.

\paragraph{Refinement and Diffusion Networks} We conduct an evaluation on the contribution of each component in \RealGDSR. Our refinement network outperforms other baselines in NMAD and MedAE with \SI{1.8}{\meter} and \SI{1.4}{\meter}, respectively (\cref{tab:comparison-metrics}, 1st row). Our diffusion network performed better than DADA because of the adjustment-step removal. Adding diffusion component the RMSE is much lower (around 20\%) but increasing both NMAD and MedAE (\cref{tab:comparison-metrics}, 3rd row), the reason is, while the diffusion network regularize the structure and removing the outliers, it also remove some of the details produced by the refinement network.


\glsresetall

\section{Conclusion}
We have presented \RealGDSRdot, a practical yet effective approach to guided super-resolution of DSMs. The approach is trying to solve the super-resolution problem in two steps: local refinement and an edge-enhancing diffusion. We highlight that the combination of CNN-based network and diffusion process bring the best of both worlds. Moreover, the local refinement network follows a residual learning strategy, i.e., it is trained to refine an imperfect bicubic-upsampled \gls{gl:DSM} by predicting correction to the height, using both the \glspl{gl:DSM} and optical images as input. Together with the diffusion step, it can leverage information from optical images not only for the local distribution of the heights but also globally preserving the edges effectively. Our approach learns to restore substantial geometry such as sharp building lines and smooth height discontinuities. Moreover, it successfully restore missing shape details in the low-resolution \gls{gl:DSM} with information from the optical images. 

In our experiments, \RealGDSR reaches top performance and outperforms state-of-the-art networks, including GANs and hybrid methods. We also found that \RealGDSR is fairly robust in terms of generalization. It can generate high-resolution \glspl{gl:DSM} of unseen test samples. We believe that if more diverse training data can be gathered from a more extensive and representative cross-section of cities, our approach can be applied universally across various regions. On a conceptual level, we hope that our work motivates further research on \gls{gl:DSM} super-resolution.

\section*{ACKNOWLEDGEMENTS}\label{ACKNOWLEDGEMENTS}

Special thanks are given to the GAF AG for the for the provision of the Cartosat-1 data. 

\bibliography{bibliography}


\end{document}